\documentclass[pra,preprint]{revtex4}
\usepackage{epsfig,amssymb, amsmath}
\usepackage{amsxtra}
\usepackage{graphicx}
\usepackage{upgreek}

\begin{document}
\title{First-Photon Target Detection:  Beating Nair's Pure-Loss Performance Limit}
\date{\today}
\author{Jeffrey H. Shapiro}\email{jhs@mit.edu}
\affiliation{Research Laboratory of Electronics, Massachusetts Institute of Technology, Cambridge, Massachusetts 02139, USA}
  
\begin{abstract} 
In 2011, Nair published a no-go theorem for quantum radar target detection~[Phys. Rev. A {\bf 84}, 032312 (2011)].  He showed, under fairly general assumptions, that a coherent-state radar's error probability was within a factor of two of the best possible quantum performance for the pure-loss (no background radiation) channel whose roundtrip radar-to-target-to-radar transmissivity $\kappa$ satisfies $\kappa \ll 1$. We introduce first-photon radars (FPRs) to circumvent and beat Nair's performance limit. FPRs transmit a periodic sequence of pulses, each containing $N_S$ photons on average, and perform ideal direct detection (photon counting at unit quantum efficiency and no dark counts) on the returned radiation from each transmission until at least one photon has been detected or a pre-set maximum of $M$ pulses has been transmitted.  They decide a target is present if and only if they detect one or more photons.  We consider both quantum (each transmitted pulse is a number state) and classical (each transmitted pulse is a coherent state) FPRs, and we show that their error-probability exponents are nearly identical when $\kappa \ll 1$.  With the additional assumption that $\kappa N_S \ll 1$, we find that their advantage in error-probability exponent over Nair's performance limit grows to 3\,dB as $M \rightarrow \infty$.  However, because FPRs' pulse-repetition period must exceed the radar-to-target-to-radar propagation delay, their use in standoff sensing of moving targets will likely employ $\kappa N_S \sim 1$ and $M \sim 10$ and achieve $\sim$2\,dB advantage.  Our work constitutes a new no-go theorem for quantum radar target detection.    
  
\end{abstract} 

\maketitle

\emph{Introduction}.---Quantum radars, as opposed to classical radars, have performance that cannot be properly quantified unless the electromagnetic fields involved are treated quantum mechanically.  In particular, classical radars radiate coherent states, or they radiate random mixtures of coherent states that are not entangled with the radar apparatus~\cite{Shapiro2020}.  Interest in quantum radars has been growing over the last decade owing to the possibility of their outperforming the best classical radars of the same transmitted energy, see Refs.~\cite{Shapiro2020,Torrome2006,Lanzagorta2012,Pirandola2018,Sorelli2022} for recent reviews and the references therein for additional details.  

Thus far, the sought after better-than-best-classical performance has only been predicted for quantum illumination radars~\cite{Tan2008,Barzanjeh2015} operating in lossy scenarios with high-brightness---many photons per mode---background radiation.  High-brightness noise prevails at microwave wavelengths and there quantum illumination offers, in principle, a 6\,dB error-exponent advantage for target detection. Unfortunately, analyses to date indicate the prospects for making microwave quantum illumination a reality are dim at best~\cite{Shapiro2020,Pirandola2018,Sorelli2022,Jonsson2021}.  

At optical wavelengths, low-brightness background radiation is the norm, e.g., at 1.55\,$\upmu$m the daytime sky has $N_B\sim 10^{-6}$\,photons/mode of background radiation~\cite{Shapiro2005}. Furthermore, nighttime $N_B$ values at that wavelength are about four orders of magnitude lower~\cite{Kopeika1970}, and 300\,K thermal radiation has $N_B\sim 10^{-14}$\,photons/mode~\cite{footnote1}.  Two recent table-top optical experiments have demonstrated quantum radars that outperformed chosen classical competitors in low-brightness background~\cite{England2019,Liu2020}.  That said, neither of those quantum radars had performance exceeding the optimum classical radar of the same transmitted energy, i.e., their chosen classical-competitor radars were distinctly suboptimum~\cite{Shapiro2020,Sorelli2022}.    
 
The challenge of finding an optical-wavelength quantum radar with better-than-best-classical performance can be appreciated from how good classical performance can be for very lossy, low-brightness noisy channels.  Here, for a coherent-state (laser) radar, the Chernoff bound on the error probability for deciding between equally-likely target absence or presence is $\Pr_{\rm CS}(e) \le e^{-\kappa N_T(\sqrt{N_B+1}-\sqrt{N_B})^2}/2 \approx e^{-\kappa N_T(1-2\sqrt{N_B})}/2$, where $0 < \kappa \ll 1$ is the roundtrip radar-to-target-to-radar transmissivity, $N_T$ is the average transmitted photon number, $N_B$ is the average received background photon number per mode, and the approximation assumes $N_B \ll 1$~\cite{Tan2008}.  For $\kappa N_T \gg 1$, this Chernoff bound's error exponent, $\kappa N_T(1-2\sqrt{N_B})$, is within $-10\log_{10}(1-2\sqrt{N_B}) \ll 1$\,dB of the Helstrom-limit error exponent, $\kappa N_T$, for a coherent-state radar operating over the pure-loss ($N_B=0$) channel with an optimum receiver~\cite{Helstrom1967}.  It would seem, therefore, that Nair~\cite{Nair2011} provided the definitive no-go result with respect to a significant better-than-best-classical performance at optical wavelengths when he showed, for equally-likely target absence or presence and a pure-loss channel, that \emph{all} radars satisfying rather general conditions had $\Pr(e) \ge (1-\sqrt{1-(1-\kappa)^{N_T}})/2 \approx e^{-\kappa N_T}/4$, where the approximation assumes $\kappa \ll 1$ and $\kappa N_T \gg 1$.  

This paper will introduce quantum (number state) and classical (coherent state) radars for the $\kappa \ll 1$ pure-loss channel that break Nair's performance limit on error exponent.  We call these new radars ``first-photon radars'', because they transmit a periodic sequence of pulses and perform ideal direct detection (photon counting with unit quantum efficiency and no dark counts) on the returned radiation from each transmission until at least one photon has been detected or a pre-set maximum of $M$ pulses has been transmitted.  They decide a target is present if and only if they detect one or more photons.  Our first-photon radars (FPRs) were inspired by our earlier work on first-photon imaging~\cite{Kirmani2014}, which used a similar laser-light (coherent-state) strategy to achieve photon-efficient---1 detected photon per pixel---3D active imaging.  For our FPRs we show that their pure-loss channel performance is virtually identical when $\kappa \ll 1$.  With the additional assumption of $\kappa N_S \ll 1$, we find that our FPRs' error-exponent advantage over Nair's performance limit grows to 3\,dB as $M \rightarrow \infty$.  However, because FPRs' pulse-repetition period must exceed the radar-to-target-to-radar propagation delay, their use in standoff sensing of moving targets will likely employ $\kappa N_S \sim 1$ and $M \sim 10$ and achieve $\sim$2\,dB advantage.  Our work constitutes a new no-go theorem for quantum radar target detection.    

The rest of this paper is organized as follows.  We begin by specifying the single-mode model for FPR operation.  Next, we analyze the single-mode model for number-state FPR, showing that it \emph{does} realize the aforementioned advantage over Nair's performance limit.  After that, we show that essentially identical performance can be obtained from a coherent-state FPR.  Finally, we discuss pertinent issues including:  (1) how these radars circumvent Nair's performance limit; (2) to what extent we can relax our initial assumptions to make coherent-state FPR practical for realistic radar scenarios; and (3) what broader lessons can be gleaned from our work.

\emph{Single-mode FPRs}.---Throughout the next two sections, we  limit our attention to single-mode FPR operation over a pure-loss channel with $\kappa \ll 1$ as described here.  First, we assume that our FPRs' goal is to distinguish the equally-likely absence (hypothesis $H_0$) or presence (hypothesis $H_1$) of an unresolved target within a particular azimuth-elevation-range-Doppler-polarization resolution bin, as is the case for Tan~\emph{et al}.'s quantum illumination~\cite{Tan2008}.  Next, we assume that the radar transmits a periodic sequence of single-mode signal pulses and does ideal direct detection (photon counting with unit quantum efficiency and no dark counts) on the radiation returned from the resolution bin of interest until either one or more photons have been detected---in which case target presence is declared---or $M$ pulses have been transmitted and target absence is declared.  To be sure that any detected photon came from the resolution bin of interest, we assume the FPRs' pulse-repetition period obeys $T \ge 2R/c$, where $R$ is that bin's range and $c$ is light speed.  Under the preceding conditions---which will be relaxed in the paper's final section---we have that the $m$th return pulse's photon annihilation operator can be taken to be 
\begin{equation}
\hat{a}_{R_m} = \left\{\begin{array}{ll}
\hat{a}_{B_m},& \mbox{$H_0$ true}\\[.05in]
\sqrt{\kappa}\,\hat{a}_{S_m} + \sqrt{1-\kappa}\,\hat{a}_{B_m},& \mbox{$H_1$ true.} \end{array}
\right.
\label{returnmodel}
\end{equation}  
Here, $\hat{a}_{S_m}$ and $\hat{a}_{B_m}$ are, respectively, the annihilation operators for the transmitted and background modes with the latter being in its vacuum state.  Note that, as in Tan~\emph{et al}.~\cite{Tan2008}, we have omitted any target-related phase shift in the target-present hypothesis.  This too will be relaxed later.  

\emph{Number-state FPR}.---A number-state FPR transmits a sequence of $N_S$-photon number states until either one or more photons have been detected or $M$ pulses have been transmitted.  We are interested in quantifying its performance by determining: its false-alarm probability $P_F \equiv \Pr(\mbox{decide $H_1$}\mid \mbox{$H_0$ true})$; its miss probability $P_M \equiv \Pr(\mbox{decide $H_0$}\mid \mbox{$H_1$ true})$; its error probability $\Pr(e) = (P_F + P_M)/2$;  its average number of transmitted pulses $\bar{m} = M/2+\sum_{m=1}^M m \Pr(m\mid H_1)/2$, where the first term is due to $M$ pulses always being transmitted when the target is absent and $\Pr(m\mid H_1)$ is the conditioinal probability that $m$ pulses are transmitted given the target is present; and its average number of transmitted photons $\bar{N} = \bar{m} N_S$.  These metrics are easily obtained, as we now show.

For the pure-loss channel and ideal photon counting there will never be any photon detections when the target is absent, hence $P_F = 0$.  Likewise, $P_M = (1-\kappa)^{MN_S}$ is the probability that no photons are detected after $M$ pulses have been transmitted when the target is present.  Thus we have $\Pr(e) = (1-\kappa)^{MN_S}/2 \approx e^{-\kappa MN_S}/2$ for $\kappa \ll 1$.  Recall that Nair's limit is $\Pr(e) \ge  (1-\sqrt{1-(1-\kappa)^{N_T}})/2 \approx e^{-\kappa N_T}/4$ for $\kappa \ll 1$, where $N_T$ is the average number transmitted signal photons.  Showing that number-state FPR's error exponent exceeds the exponent from Nair's lower bound is our next task.  It should already be obvious that it will, because $MN_S$ is number-state FPR's \emph{maximum} number of transmitted signal photons, so its average number must be lower unless no photons are ever detected in less than $M$ pulses.  Hence, for the same average transmitted photon number, the number-state FPR's error exponent \emph{must} be greater than that of Nair's lower bound.  The question is by how much.  

Because the number-state FPR transmits $N_S$-photon number states, we have that
\begin{eqnarray}
\lefteqn{\Pr(m\mid H_1) = }\nonumber \\[.05in]
&&\hspace*{-0.1in}\left\{\begin{array}{ll}
[1-(1-\kappa)^{N_S}](1-\kappa)^{(m-1)N_S}, &\mbox{$m = 1,2,\ldots,M-1$}\\[.1in]
(1-\kappa)^{(M-1)N_S}, &\mbox{$m = M$,}
\end{array}\right.
\end{eqnarray}
from which we get
\begin{equation}
\bar{m} = M/2 + \frac{1-(1-\kappa)^{MN_S}}{2(1-(1-\kappa)^{N_S})} \approx M/2 + \frac{1-e^{-\kappa M N_S}}{2(1-e^{-\kappa N_S})},
\label{NSnbar}
\end{equation}
for $\kappa \ll 1$.   This result further reduces to $\bar{m} \approx M/2+1/2\kappa N_S$ when $\kappa N_S \ll 1$ and $\kappa M N_S \gg 1$.   In this asymptotic regime the number-state FPR transmits  $\bar{N} \approx MN_S/2+1/2\kappa$ photons on average to achieve $\Pr(e) \approx e^{-\kappa M N_S}/2$, whereas Nair's lower bound for $\kappa \ll 1$ is $\Pr(e) \ge e^{-\kappa N_T}/4$, where $N_T$ must be set to the number-state FPR's $MN_S$ to match their error exponents.  In the $\kappa \ll 1$, $\kappa N_S \ll 1$ regime, number-state FPR's advantage over Nair's performance limit saturates at 3\,dB 

Unfortunately, FPR operation in the $\kappa \ll 1$, $\kappa N_S \ll 1$ regime is unlikely to be of use for standoff sensing as its dwell time, $MT \ge 2MR/c$, will be too long for target ranges of interest with moving targets.  Operating with $\kappa N_S \sim 1$, however, provides a reasonable compromise between the conflicting goals of minimizing $M$ and minimizing $\bar{N}$ for a chosen $\Pr(e)$.  For example, with $\kappa = 10^{-6}$ and $\Pr(e) = 10^{-6}$, we can use $M=10$ to achieve $\bar{m} = 6.37$ and $\bar{N} = 1.79 \times 10^6$, which is 1.80\,dB \emph{lower} than the $N_T = 1.24 \times 10^7$ needed for Nair's lower bound to be $\Pr(e) \ge 10^{-6}$.  

\emph{Coherent-state FPRs}.---Present technology is still struggling to optimize the generation of single-mode single photons on demand, so the prospect of generating the $N_S = 10^6$ number state on demand seems out of the question.  Coherent states, however, with $10^6$ photons on average are easily obtained and have photon number standard deviations of $10^3$, so we might expect that a coherent-state FPR could mimic the performance we have just found for the number-state FPR.  Such is indeed the case, as we now show.

For the coherent-state FPR we have that the performance metrics considered in the previous section become:  $P_F = 0$, $P_M = e^{-\kappa M N_S}$, $\Pr(e) = e^{-\kappa M N_S}/2$, 
\begin{eqnarray}
\lefteqn{\Pr(m\mid H_1) = }\nonumber\\[.05in]
&&\left\{\begin{array}{ll}
[1-e^{-\kappa N_S}]e^{-(m-1)\kappa N_S}, &\mbox{$m = 1,2,\ldots,M-1$}\\[.1in]
e^{-\kappa (M-1)N_S}, &\mbox{$m = M$;}
\end{array}\right.
\end{eqnarray}
\begin{equation}
\bar{m} = M/2 + \frac{1-e^{-\kappa M N_S}}{2(1- e^{-\kappa N_S})},
\label{CSnbar}
\end{equation}
for $\kappa \ll 1$, which becomes $M/2 + 1/2\kappa N_S$ when $\kappa N_S \ll 1$ and $\kappa MN_S \gg 1$.  So, coherent-state FPR reproduces the number-state FPR's asymptotic performance, viz., $\bar{m} \approx M/2 + 1/2\kappa N_S$ and $\bar{N} \approx MN_S/2+ 1/2\kappa$, but, as before, much of that 3\,dB maximum advantage can be realized with $M$ values consistent with realistic dwell times for standoff sensing. 
 
Because the $\kappa \ll 1$ approximation in Eq.~\eqref{NSnbar} matches the exact result in Eq.~\eqref{CSnbar} we expect both to have the same advantage over Nair's lower bound.  Indeed, it is easily verified that for $\kappa \le 10^{-3}$, and number-state FPR error probabilities satisfying $10^{-9}\le \Pr^{\rm FPR}_{\rm NS}(e) \le 10^{-1}$, the coherent-state FPR's error probability, $\Pr^{\rm FPR}_{\rm CS}(e)$, will be no more than $1.01 \Pr^{\rm FPR}_{\rm NS}(e)$ at the same $N_S$ value.  Moreover, over most of that parameter region the two error probabilities will be considerably closer than that.    This error-probability confluence constitutes the FPR version of Nair's no-go theorem for an $N_S \gg 1$ quantum radar to have a significant better-than-best-classical performance advantage~\cite{footnote2}.  Note that despite the performance equivalence between number-state and coherent-state FPRs when $N_S \gg 1$, the latter enjoys a great advantage over the former in that parameter region.  In particular, the coherent-state FPR has a viable source for its $N_S \gg 1$ transmitter---a laser---whereas one is not available, nor even likely to ever be available, for the photon-number FPR. 

\emph{Discussion}.---By this point we have presented compelling evidence that Nair's lower bound on target-detection error probability for a radar operating over a pure-loss channel can be broken, at least in principle.  It now behooves us to explain why our results are \emph{not} in conflict with Nair's.  The answer is rather simple.  Nair assumed that the radar transmitted a pure state of deterministic time duration.  FPRs, however, transmit a random number of pulses to detect a target.  Hence, each FPR decision occurs after a random amount of transmission time.  Averaged over that randomness, FPRs transmit mixed states.  Thus FPRs' violating Nair's performance limit is not a contradiction.  

How FPRs beat Nair's performance limit also has a simple answer:  FPRs exploit vacuum-or-not decisions, which can be performed \emph{without} false-alarm errors.  Consider the likelihood-ratio test (LRT) for a single-pulse transmission by our coherent-state FPR operating over a pure-loss channel in the event that $n$ photons are detected:
\begin{equation}
\frac{\Pr(n\mid H_1)}{\Pr(n\mid H_0)} = \left\{\begin{array}{ll}
e^{-\kappa N_S},& \mbox{for $n = 0$}\\[.05in]
\infty, &\mbox{for $n \ge 1$.}\end{array}\right. \begin{array}{cc}
\mbox{\scriptsize decide $H_1$}\\ \ge \\ < \\ \mbox{\scriptsize decide $H_0$}\end{array} 1. 
\end{equation}  
This LRT is singular, i.e., it makes a vacuum-or-not decision based on ideal direct detection that achieves $P_F = 0$ and $P_M = e^{-\kappa N_S}$.  Repeated use of this test by transmission of a sequence of pulses will then drive down the multi-pulse $P_M$ below any desired value. 

Having explained why and how FPRs can beat Nair's performance limit, we turn to the task of relaxing the assumptions we made at the outset of our study.   Specifically, we ask whether FPRs can preserve most---if not all---of their performance advantage despite unknown target-return phase, unknown target-return polarization, Doppler and range uncertainties that are greater than the radar receiver's resolution limits~\cite{footnote3}, sub-unit detector quantum efficiency, low-brightness background radiation, and detector dark counts.  The answers are, for the most part, yes.  Unknown target-return phase and unknown target-return polarization pose no problems because a direct-detection receiver is insensitive to target-return phase and detects both polarizations.  For the pure-loss channel, no performance will be lost if we open up the optical bandwidth of our detection scheme to encompass the Doppler uncertainty.  Likewise, for the pure-loss channel, no performance will be lost if we look for photodetections occurring at range delays corresponding to the range-uncertainty interval, $R \in [R_{\rm min},R_{\rm max}]$, and use $T \ge 2R_{\rm max}/c$.  Sub-unit detector quantum efficiency, $\eta < 1$, will only affect performance on the pure-loss channel by changing $\kappa$ to $\kappa'\equiv \eta\kappa$ in our FPR performance analyses.  Inasmuch as superconducting nanowire single-photon detectors (SNSPDs) are commercially available with quantum efficiencies exceeding 0.85, the performance lost to this effect is hardly consequential for the $\kappa \ll 1$ values encountered in standoff sensing.  It is low-brightness background radiation and detector dark counts that pose significant problems for FPR operation, so they command discussion at length.

The presence of low-brightness background radiation and/or dark counts breaks the vacuum-or-not decision rule that underlies our FPR analysis.  In order that these effects imply a maximum factor-of-two degradation in error probability, it suffices to ensure that the false alarms they create occur with no more than the miss probabilities we calculated earlier.  Toward understanding whether and when the preceding condition might apply, let us consider the task laid out in the first three columns of Table~\ref{resolutionnumbers}.  There we assume a $\lambda = 1.55\,\upmu$m wavelength, coherent-state FPR with $D = 5\,$cm diameter optics that uses up to $M = 10$, $T= 0.1\,\upmu$s duration transform-limited pulses to distinguish the absence or presence of a 25-cm-long, quad-rotor drone at $\sim$10\,km range that is moving transversely at 10\,km/hr speed and longitudinally at 10\,km/hr speed.   The left three columns give the FPR's angle, range, and Doppler resolutions, while the right three give the angle, range, and Doppler changes over the radar's dwell time created by the drone's transverse speed, its longitudinal speed, and its $a_z = 0.98\,\mbox{m/s$^2$}$ longitudinal acceleration.  Because the drone subtends 25\,$\upmu$Rad at 10\,km range, Table~\ref{resolutionnumbers} shows that it will stay within a single angle-resolution bin over the FPR's dwell time.  Likewise, it will stay within a single range-resolution bin over the FPR's dwell time, but, taking the target's range uncertainty to be $\sim\pm0.5$\,km, the FPR will need to look for photodetections over $B_r \sim 70$ range bins.  Similarly, although the drone will stay within one Doppler-resolution bin over the FPR's dwell time, taking the target's Doppler uncertainty to be $\sim\pm 0.5$\,GHz, the FPR will need to measure photodetections over $B_d \sim 100$ Doppler bins.  All told, the system whose parameters are given in Table~\ref{resolutionnumbers} might therefore need to look for photodetections over $B = B_rB_d \sim 7000$ angle-range-Doppler resolution bins for each transmitted pulse.  Daytime $N_B \sim 10^{-6}$ would lead to $P_F = 1-e^{-BMN_B}\sim 0.1$, which is doubtless unacceptable.  Nighttime $N_B \sim 10^{-10}$, however, would give $P_F \sim 10^{-5}$, implying that our $M=10$ coherent-state FSR could approach pure-loss performance at $\Pr(e) \sim 10^{-4}$ in the absence of dark counts.  Commercially available SNSPDs have ${\rm DCR} < 100\,$cps dark-count rates, implying they cause false alarms with $P_F = 1-e^{-\text{DCR}MB_rT} \sim 0.01$, ruling out good FPR performance without substantially smaller range uncertainties or lower DCRs.     
\begin{center}
\begin{table}[h]
\begin{tabular}{|c|c|c||c|c|c|}\hline
parameter & symbol & value & parameter & symbol & value  \\ \hline
angle resolution & $\theta_{\rm res} = 1.22 \lambda/D$ & $38\,\upmu$Rad & angle change in dwell time& $\Delta\theta = v_t T_d$ & $0.19\,\upmu$Rad\\ 
range resolution & $R_{\rm res}= cT/2$ &15\,m &range change in dwell time & $\Delta R \equiv v_zT_d$ & 6.7\,m \\
Doppler resolution & $f_{\rm res}= 1/T$ & 10\,MHz & Doppler change in dwell time  & $2a_z T_d/\lambda$ & 0.42\,kHz \\ \hline
\end{tabular}
\caption{Coherent-state, $M = 10$, FPR example for evaluation of background radiation and dark-count effects on the detection of a 25-cm-long, slow-moving, quad-rotor drone at $R\sim 10\,$km range that is moving transversely at $v_t = 10\,$km/hr speed and longitudinally at $v_z = 10\,$km/hr speed.  Left three columns:  angle, range, and Doppler resolutions for a $\lambda = 1.55\,\upmu$m wavelength, $D = 5\,$cm optics diameter, $T = 0.1\,\upmu$s pulse duration (transform limited), $T_r = 2R/c = 67\,\upmu$s pulse-repetition interval, and $T_d = 2 M R/c = 0.67\,$ms dwell-time radar.   The Doppler shift for this $v_z$ is $f_d \equiv  2v_z/\lambda = 13\,$GHz. Right three columns:  angle, range, and Doppler changes caused by $v_t$, $v_z$ and longitudinal acceleration $a_z = 0.1\,g$ where $g=9.8\,\mbox{m/s$^2$}$ is the gravitational acceleration. \label{resolutionnumbers}} 
\end{table}
\end{center}

Some final remarks about the vacuum-or-not sequential measurement that enables our FPR's advantage over Nair's performance limit are now in order.  In 1973, Kennedy~\cite{Kennedy1973} described a near-optimum receiver for binary phase-shift keyed coherent-state communication over the pure-loss channel.  It used a vacuum-or-not measurement that, unlike conventional receivers, achieved the Helstrom-limit error  exponent.  Later, Wilde \emph{et al}.~\cite{Wilde2012} showed that a non-destructive vacuum-or-not measurement could be used to achieve the pure-loss channel's Holevo capacity for classical information transmission.  Recently, Jagannathan~\emph{et al}.~\cite{Jagannathan2022} showed---through theory and experiment---that repeated use of the Kennedy receiver's vacuum-or-not measurement enabled the Chernoff-bound exponent for multicopy discrimination between a coherent state and a thermal state to be realized.  Collectively, those studies and our work here should motivate additional efforts to exploit vacuum-or-not techniques.  Also, because FPRs' decision rule is a special case of Wald's sequential probability-ratio test (SPRT)~\cite{Wald1945}, the use of SPRTs in other quantum radars, e.g., those operating in high-brightness background radiation, is warranted.  

This research was supported by the MITRE Corporation's Quantum Moonshot Program. 
The author acknowledges R. Nair for pointing out an egregious blunder in version~1 of this manuscript.


\begin{thebibliography}{2}
\bibitem{footnote1} For $T_B\,$K thermal radiation we have $N_B= 1/(e^{hc/\lambda kT_B} -1)$ at wavelength $\lambda$, where $h$ is Planck's constant, $c$ is light speed, and $k$ is Boltzmann's constant.
\bibitem{footnote2}The optimum state, $|\psi\rangle$, for an FPR  minimizes the probability of zero target-return counts, i.e., $\sum_{n=0}^\infty (1-\kappa)^n |\psi_n|^2$, where $\sum_{n=0}^\infty \psi_n|n\rangle$ is its number-state representation and $\sum_{n=1}^\infty n|\psi_n|^2 = N_S$. The convexity of $(1-\kappa)^x$ implies that $|\psi\rangle = \sqrt{p}\,|\underline{N_S}\rangle + \sqrt{1-p}\,|\overline{N_S}\rangle$  where $0\le p\le 1$ is such that $p\underline{N_S} + (1-p)\overline{N_S} = N_S$, with $\underline{N_S}$ being the largest integer $\le N_S$ and $\overline{N_S}$ being the smallest integer $\ge N_S$.  For $N_S \gg 1$, the number-state $|N_S\rangle$ is thus essentially optimum.
\bibitem{footnote3}We will not consider azimuth-elevation uncertainties that exceed the radar receiver's resolution limit because we are continuing to assume that the target is unresolved.  With a coherent-state FPR we can interrogate $K$ azimuth-elevation resolution bins simultaneously by floodlighting that angle uncertainty region and then use a $K$-element photon-counting array in the receiver.  Assuming the same photon flux is transmitted as in the single-bin case, $K$-bin floodlighting reduces $\kappa$ by a factor of $1/K$ for  the FPR and the optimum quantum radar, implying that FPRs continues to beat Nair's performance limit.
\end{thebibliography}
\end{document}